\documentclass[aps,twocolumn,showpacs]{revtex4}
\usepackage{graphicx}

\begin{document}

\title{Corrugated waveguide under scaling investigation}

\author{Edson D.\ Leonel}

\affiliation{Departamento de Estat\'{\i}stica, Matem\'atica
Aplicada e Computa\c c\~ao -- Instituto de Geoci\^encias e
Ci\^encias Exatas -- Universidade Estadual Paulista\\
Av.24A, 1515 -- Bela Vista -- CEP: 13506-700 -- Rio Claro -- SP --
Brazil}

\date{\today} \widetext


\begin{abstract}
Some scaling properties for classical light ray dynamics inside a
periodically corrugated waveguide are studied by use of a
simplified two-dimensional nonlinear area-preserving map. It is
shown that the phase space is mixed. The chaotic sea is
characterized using scaling arguments revealing critical exponents
connected by an analytic relationship. The formalism is widely
applicable to systems with mixed phase space, and especially in
studies of the transition from integrability to non-integrability,
including that in classical billiard problems.
\end{abstract}
\maketitle

Interest in the problem of {\it guiding a light ray} inside a
periodically corrugated boundary has increased in recent years,
partly because the topic is applicable in so many different fields
of science. Applications involve e.g.\ ray chaos in underwater
acoustics \cite{ref1,ref2,ref3}, scattering of a quantum particle
in a rippled waveguide \cite{ref4}, quantized ballistic
conductance in a periodically modulated quantum channel
\cite{ref5}, quantum transport in ballistic cavities \cite{ref6},
transport through a finite ${\rm GaAs/Al_xGa_{1-x}As}$
heterostructure \cite{ref8}, comparisons of classical v.\ quantum
behavior in periodic mesoscopic systems \cite{ref9}, and anomalous
wave transmittance in the stop band of a corrugated parallel-plane
waveguide \cite{ref11}.

There are many different ways of describing problems involving
waveguides. One of them is the well known billiard description,
and such a procedure is used in the present paper. A billiard
problem consists of a system in which a point particle moves
freely inside a bounded region and suffers specular reflections
with the boundaries. Generally, the dynamics is described using
the formalism of discrete maps. Depending on the combinations of
control parameter values, as well as on the initial conditions,
the phase spaces for such mappings fall into three distinct
classes, namely: (i) regular; (ii) ergodic; or (iii) mixed.
Generally, the integrability of the regular cases is related to
angular momentum conservation, the static circular billiard being
a typical example. On the other hand, for completely ergodic
billiards, only chaotic and unstable periodic orbits are present
in the dynamics. Two examples of case (ii) are the Bunimovich
stadium \cite{ref12} and the Sinai billiard \cite{ref14}. For
these two systems, the time evolution of a single initial
condition, for the appropriate combinations of control parameters,
is enough to fill the whole phase space, ergodically. Finally,
there are many billiards of mixed phase space structure
\cite{ref15,ref17,ref18,ref19,add1,add3}, having a range of
control parameters whose physical significances differ. Depending
on the combination of both initial conditions and control
parameters, the phase space presents a very rich structure
containing invariant spanning curves (sometimes known as invariant
tori), Kolmogorov-Arnold-Moser (KAM) islands, and chaotic seas.

In this Letter, I use scaling arguments to describe the behavior
of the variance of the average reflection angle, within the
chaotic sea, for a classical light ray undergoing specular
reflections inside a periodically corrugated waveguide. The model
consists of a classical light ray that is specularly reflected
between a corrugated surface given by $y=y_0+d\cos(k x)$ and a
flat plane surface at $y=0$. The term $y_0$ denotes the average
distance between the corrugated and flat surfaces, $d$ is the
amplitude of the corrugation and $k$ is the wave number. The
dynamical variables used in the description of the problem are the
angle $\theta$ of the ray's trajectory measured from the positive
horizontal axis, and the corresponding value of the $x$ coordinate
at the instant of reflection. Moreover, the mapping is iterated
when the light {\it hits} the surface $y=0$; thus multiple
reflections with the corrugated surface can be neglected.

\begin{figure}[htb]
\vspace*{-0.8cm}
\centerline{\includegraphics[width=0.85\linewidth]{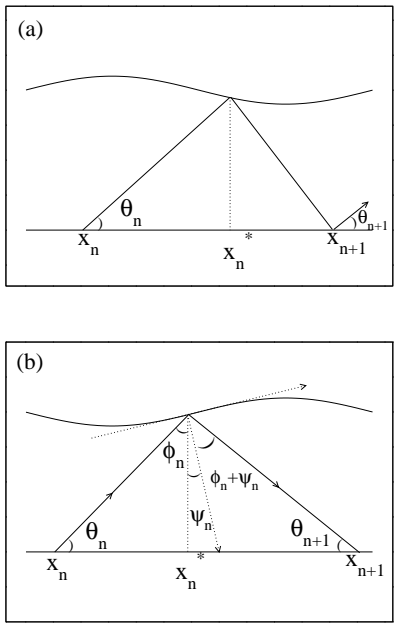}}
\caption{(a) Reflection from the corrugated surface of a light ray
coming from the flat surface at $y=0$. (b) Details of the
trajectory before and after a collision with the corrugated
surface.} \label{Fig1}
\end{figure}
The problem lies in obtaining the map
$T(x_n,\theta_n)=(x_{n+1},\theta_{n+1})$, given the initial
conditions $(x_n,\theta_n)$ shown in Fig.\ \ref{Fig1}(a). From the
geometrical considerations illustrated in the figure, it is easy
to see that
$x_n^*-x_n=(y_0+d\cos(k x_n^*))/\theta_n$,
and similarly that
\begin{equation}
x_{n+1}-x_n^*=(y_0+d\cos(k x_n^*))/\theta_{n+1}~.
\label{eq2}
\end{equation}
The term $x_n^*$ gives the location of the collision on the
corrugated surface. The angle $\theta_n$ is written as
\begin{equation}
\theta_{n+1}=\theta_n-2\psi_n~,
\label{eq3}
\end{equation}
where $\tan(\psi_n(x))=dy(x)/dx=-dk\sin(k x_n^*)$ gives the slope
of the surface at $x=x_n^*$. Equations (\ref{eq2}) and (\ref{eq3})
correspond to the exact mapping. However in this paper we make the
following approximations: (a) we assume that $d/y_0\ll 1$, so that
$y_0+d\cos(k x_n^*)\cong y_0$; (b) in the same limit, we also
assume that $\tan(\psi_n)\cong\psi_n$. The condition $d/y_0\ll 1$
holds in practice for studies of wave propagation in a corrugated
waveguide \cite{ref3}. It also applies to the investigation of
transport in mesoscopic channels \cite{ref9} and can even be used
in the characterization of wave propagation in deep ocean
\cite{ref1,ref2}. In the description of the mapping, it is
convenient to use dimensionless variables, $\delta=d/y_0$,
$\gamma_n=\theta_n/k$ and $X_n=k x_n/y_0$. Thus, the simplified
two-dimensional nonlinear mapping is given by
\begin{equation}
T:\left\{\begin{array}{ll}
X_{n+1}=X_n+\left[{{1}\over{\gamma_n}}+{{1}\over{\gamma_{n+1}}}\right]~~
{\rm mod (2\pi)}\\
\gamma_{n+1}=\gamma_n+2\delta\sin\left(X_n+{{1}\over{\gamma_n}}
\right)~~\\
\end{array}
\right.~.
\label{eq4}
\end{equation}
The determinant of the Jacobian matrix for this mapping is unity,
so that it is area-preserving. The mapping (\ref{eq4}) is
described by a single effective control parameter, $\delta$. If
$\delta=0$, the system is integrable while for $\delta\ne 0$, the
phase space is mixed, containing both chaos and regularity (fixed
points and quasiperiodic behavior). The system thus experiences an
abrupt transition from integrability to non-integrability when
going from $\delta=0$ to $\delta\ne 0$. I shall investigate some
dynamical properties for very small values of $\delta$ (which
meets the initial approximation $d/y_0\ll 1$) near the transition,
using scaling arguments. There is a wide range of systems
exhibiting transitions from integrability to non-integrability,
and the formalism used in the present model can be in principle be
extended to them, in particular to identify classes of
universality.

\begin{figure}[t]
\centerline{\includegraphics[width=0.9\linewidth]{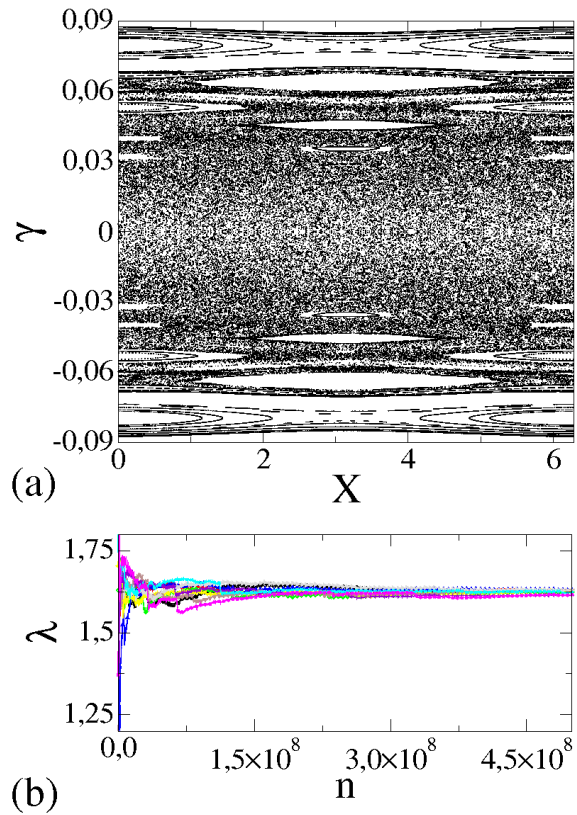}}
\caption{(Color online) (a) Phase space for the mapping
(\ref{eq4}) with the control parameter $\delta=10^{-3}$. (b)
Positive Lyapunov exponent obtained via the triangularization
algorithm.} \label{Fig2}
\end{figure}

Fig.\ \ref{Fig2}(a) shows the phase space generated by iterating
the map (\ref{eq4}) for the control parameter $\delta=10^{-3}$. It
is easy to see that the phase space is of mixed form, including
invariant spanning curves, KAM islands and a large chaotic sea for
both positive and negative values of the variable $\gamma$. The
invariant spanning curves limit the size of the chaotic sea and
correspondingly yield a limited range of possible values for the
chaotic orbits. Fig. \ref{Fig2}(b) shows positive Lyapunov
exponents obtained for the chaotic sea using the triangularization
algorithm \cite{ref24}, for the control parameter
$\delta=10^{-3}$. An ensemble of $10$ different initial conditions
was iterated $5\times 10^8$ times. The initial conditions
considered were $X_0=\delta$ and 10 different values of $\gamma_0$
uniformly distributed in the interval $\gamma\in[0,2\pi)$. The
average value obtained was ${\bar{\lambda}}=1.624(7)$, where 0.007
is the standard deviation obtained from the 10 samples.

The existence of invariant spanning curves limiting the size of
the chaotic sea confers an interesting property on the chaotic
orbits. From an initial condition in the region of the chaotic
sea, the system wanders within the entire accessible region but it
is always confined between two invariant spanning curves. The
location of the first positive and negative invariant spanning
curves depends on the value of the control parameter. As a
consequence, the ``amplitude'' of a chaotic time series is also
dependent on the control parameter, so that the average value
tends only gradually towards a regime of convergence. With this
property in mind, I now explore the behavior of the deviation of
the average value for the angle $\gamma$, or {\it roughness}, an
observable defined as
\begin{equation}
\omega(n,\delta)={{1}\over{M}}\sum_{i=1}^{M}\sqrt{\bar{\gamma_i^2}(n,
\delta)-
{\bar{\gamma_i}}^2(n,\delta)}~,
\label{eq5}
\end{equation}
where
\begin{equation}
\bar{\gamma}(n,\delta)={{1}\over{n}}\sum_{i=0}^n\gamma_i~.
\label{eq6}
\end{equation}

\begin{figure}[htb]
\centerline{\includegraphics[width=0.9\linewidth]{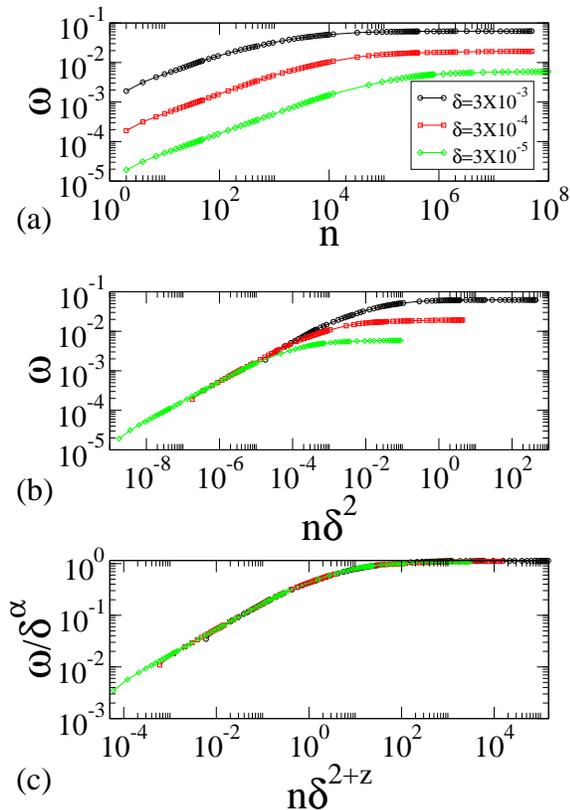}}
\caption{(Color online)  (a) Roughness $\omega$ as a function of
the iteration number $n$ for three different control parameters.
(b) After a suitable change of variables, all three curves start
by growing together. (c) Their collapse onto a single universal
plot.} \label{Fig3}
\end{figure}

\noindent Equation (\ref{eq5}) was iterated using an ensemble of
$M=5\times 10^3$ different initial conditions. The variable
$\gamma_0$ was kept fixed at $\gamma_0=10^{-2}\delta$ while the
$5000$ values of $X_0$ were uniformely distributed along
$X_0\in[0,2\pi)$. Fig.\ \ref{Fig3}(a) shows the behavior of three
different values of the roughness for different control
parameters.
One can observe that, as the number of iterations increases, the
roughness grows at first, but then suddenly bends over towards a
regime of saturation. The
\begin{figure}[htb]
\centerline{\includegraphics[width=0.9\linewidth]{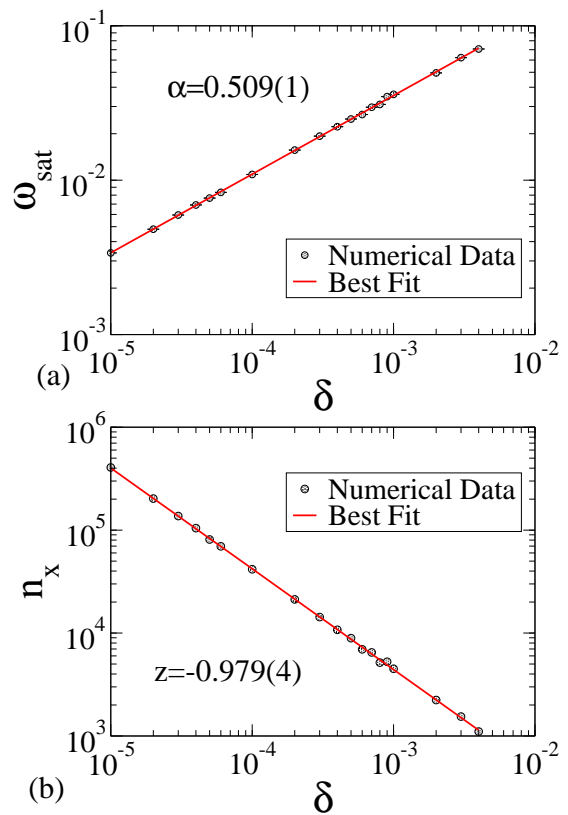}}
\caption{(Color online) Plots of (a) $\omega_{\rm sat}$ and (b)
the crossover iteration number $n_x$ as function of the control
parameter $\delta$.} \label{Fig4}
\end{figure}
changeover from growth to saturation is marked by a characteristic
crossover iteration number denoted as $n_x$. Note that different
values of the control parameter generate different roughness
curves for small iteration numbers $n$. Applying the
transformation $n\rightarrow n\delta^2$ yields a coalescence of
all the curves at small iteration number, as can be seen in Fig.
\ref{Fig3}(b). Based on the behavior of Fig. \ref{Fig3}(a), the
following three scaling hypotheses are proposed: (i) that for
small iteration numbers, say $n\ll n_x$, the roughness behaves
according to the power law
\begin{equation}
\omega(n\delta^2,\delta)\propto(n\delta^2)^{\beta}~,
\label{eq7}
\end{equation}
where we refer to the critical exponent $\beta$ as the {\it growth
exponent}; (ii) for large enough iteration number, $n\gg n_x$, the
roughness approaches a regime of saturation marked by a constant
``plateau'' given by
\begin{equation}
\omega_{\rm sat}\propto\delta^{\alpha}~,
\label{eq8}
\end{equation}
where we call the exponent $\alpha$ the {\it roughening exponent};
finally (iii) the number of iterations that characterizes the
crossover, i.e.\ that marks the change from growth to saturation
is written as
\begin{equation}
n_x\propto \delta^z~.
\label{eq9}
\end{equation}
The exponent $z$ is a {\it dynamical exponent}. These three
scaling hypothesis are extensions of the formalism used in surface
science (see e.g.\ Ref. \cite{add2}). Based on these three initial
suppositions, the roughness can now be described in terms of a
scaling function of the type
\begin{equation}
\omega(n\delta^2,\delta)=l\omega(l^an\delta^2,l^b\delta)~,
\label{eq10}
\end{equation}
where $l$ is a scaling factor, and $a$ and $b$ are scaling
exponents. Moreover, the exponents $a$ and $b$ must be related to
the critical exponents $\alpha$, $\beta$ and $z$. Because $l$ is a
scaling factor, we can specify that $l=(n\delta^2)^{(-1/a)}$ and
rewrite Eq.\ (\ref{eq10}) as
\begin{equation}
\omega(n\delta^2)=(n\delta^2)^{(-1/a)}\omega_1([n\delta^2]^{(-b/a)}
\delta)~,
\label{eq11}
\end{equation}
where the function $\omega_1=\omega(1,[n\delta^2]^{-b/a}\delta)$
is assumed to be constant in the limit of $n\ll n_x$. Comparison
of Eqs. (\ref{eq11}) and (\ref{eq7}) allows one to conclude
immediately that $\beta=-1/a$. Choosing now that
$l=\delta^{-1/b}$, Eq. (\ref{eq10}) is given by
\begin{equation}
\omega(n\delta^2,\delta)=\delta^{-1/b}\omega_2(\delta^{-(a/b)}
n\delta^2)~,
\label{eq12}
\end{equation}
where the function $\omega_2=\omega(\delta^{-(a/b)}n\delta^2,1)$
is supposed constant for $n\gg n_x$. Comparing Eqs.\ (\ref{eq12})
and (\ref{eq8}), we find that $\alpha=-1/b$. Given the two
different expressions for the scaling factor $l$, one must
conclude that the relation between the critical exponents takes
the form
\begin{equation}
z={{\alpha}\over{\beta}}-2~.
\label{eq13}
\end{equation}
Note that the scaling exponents are all determined if the critical
exponents $\alpha$ and $\beta$ can be obtained numerically. Fig.\
\ref{Fig4}(a) and (b) show the behavior of $\omega_{\rm sat}$ and
$n_x$ respectively as functions of $\delta$. The saturation values
were obtained by extrapolation because, even after almost
$10^3n_x$ iterations, the roughness is still not approaching its
saturation value. Fitting a power law to the results of Figs.
\ref{Fig4}(a) and (b), it was found that $\alpha=0.509(1)$,
$\beta=0.4997(8)$ and $z=-0.979(4)$. The dynamical exponent $z$
can also be obtained from Eq.\ (\ref{eq13}), by evaluation of the
numerical values of $\alpha$ and $\beta$, yielding $z=-0.9814(3)$.
This result is in excellent agreement with the numerical result
obtained in Fig. \ref{Fig4}(b).

With the values of the critical exponents now obtained, the
scaling hypotheses can be verified. Fig. \ref{Fig3}(c) illustrates
the collapse of the three different roughness curves generated
from different values of the control parameter onto a single
universal roughness plot. Such a collapse confirms that the
scaling suppositions are indeed correct despite the three decades
used in the control parameter $\delta$. In the construction of all
the roughness curves, the value of the dynamical variable
$\gamma_0$ was set at $\gamma_0=10^{-2}\delta$. If one increases
the initial value of $\gamma_0$, the scaling hypotheses will
remain valid, but an extra scaling time (say $n_x^{\prime}$)
appears. Even considering the new extra crossover iteration
number, scaling exponents can still be obtained \cite{ref22}.

In summary, the problem of light ray reflection in a periodically
corrugated waveguide has been addressed. The chaotic sea confined
between the first two invariant spanning curves (the positive and
negative ones) was characterized using scaling arguments. In
particular, the deviation of the average value of $\gamma$ was
obtained as functions of both the control parameter $\delta$ and
the iteration number $n$. Three scaling hypotheses were proposed,
and then confirmed by the perfect collapse of all curves onto a
single universal plot, confirming that the system experiences a
transition from integrability to non-integrability. Finally, the
obtained critical exponents $\alpha$, $\beta$ and $z$ allow us to
conclude that the present corrugated waveguide model belongs to the
same class of universality of the one-dimensional bouncing ball model.

E.D.L. thanks Prof. J. K. L. da Silva and Prof. P.V.E. McClintock for
helpful discussions. Support from CNPq, FAPESP and FUNDUNESP,
Brazilian agencies is gratefully acknowledged.

\end{document}